\documentclass[fleqn,usenatbib]{mnras}

\usepackage{newtxtext,newtxmath}

\usepackage[T1]{fontenc}

\DeclareRobustCommand{\VAN}[3]{#2}
\let\VANthebibliography\thebibliography
\def\thebibliography{\DeclareRobustCommand{\VAN}[3]{##3}\VANthebibliography}

\usepackage{graphicx}	
\usepackage{amsmath}	

\newcommand{\msun}[0]{\mathrm{M}_\odot}

\title[Nuclear Star Clusters in LYRA]{LYRA: The formation and growth of nuclear star clusters in dwarf galaxies}

\author[J. Sureda et al.]{
Joaquin Sureda,$^{1,2}$\thanks{E-mail: joaquin.m.sureda@durham.ac.uk},
Azadeh Fattahi$^{1,3}$, 
Sownak Bose$^{1}$,
Mariya Lyubenova$^{2}$,
Jessica E. Doppel$^{1,4}$,
\newauthor
Thales Gutcke$^{5}$,
Katja Fahrion$^{6}$,
Rüdiger Pakmor$^{7}$
\\
$^{1}$Institute for Computational Cosmology, Department of Physics, Durham University, South Road, Durham DH1 3LE, UK\\
$^{2}$European Southern Observatory, Karl-Schwarzschild-Straße 2, 85748 Garching bei München, Germany.\\
$^{3}$The Oskar Klein Centre, Department of Physics, Stockholm University, Albanova University Center, 106 91 Stockholm, Sweden\\
$^{4}$Centre for Extragalactic Astronomy, Department of Physics, Durham University, Durham DH1 3LE, UK\\
$^{5}$Institute for Astronomy, University of Hawaii, 2680 Woodlawn Drive, Honolulu, HI 96822, USA\\
$^{6}$Department of Astrophysics, University of Vienna, T\"{u}rkenschanzstra{\ss}e 17, 1180 Wien, Austria\\
$^{7}$Max-Planck-Institut für Astrophysik, Karl-Schwarzschild-Str. 1, D-85748, Garching, Germany
}

\date{Accepted XXX. Received YYY; in original form ZZZ}

\pubyear{\the\year{}}

\begin{document}
\label{firstpage}
\pagerange{\pageref{firstpage}--\pageref{lastpage}}
\maketitle

\begin{abstract}
Nuclear star clusters (NSCs) are dense massive star clusters ubiquitous in galactic centres across various mass regimes, present in up to $\sim30$ per cent of dwarf galaxies with $M_\star<10^{7}\,\msun$. Although their main growth pathways are well studied, there is still ongoing debate regarding their formation. We use the LYRA cosmological hydrodynamical simulations to study the formation and growth of NSCs in a sample of six dwarf galaxies, where we identify NSCs with masses $M_\mathrm{NSC} \sim 10^4 - 10^6 \, \msun$ in all galaxies in our sample. The NSCs in our sample emerge at early times as a compact component, with the host galaxy growing around them. This is consistent with a scenario in which NSCs are among the earliest structures to form in a galaxy. We characterise the origin of the stars that constitute the NSCs and explore the properties of the stellar populations in each of these channels. We find that most of the NSC mass originates from in-situ star formation within the NSCs themselves. While this might appear at odds with current observations, the bulk of this star formation occurs at early times, resulting in a significant population of old and metal-poor stars, in agreement with an early NSC formation. We also identify galaxy-galaxy mergers as a significant contributor to the NSC mass, with usually one merger dominating this contribution. Our results highlight the complexity in the growth history of NSCs in dwarf galaxies and highlight the importance of considering additional growth channels for NSCs in dwarfs galaxies.
\end{abstract}




\section{Introduction}

Nuclear star clusters (NSC) are the densest stellar structures in the Universe and are found at the centres of most galaxies, ranging from large spirals and ellipticals to dwarf galaxies. Despite their relatively small size, NSCs are useful probes of galaxy evolution. For instance, they are connected to massive black hole (MBH) growth \citep{Antonini:2013, Antonini:2015, Partmann:2025} and can even serve as unique probes of central dark matter (DM) density profiles \citep{Herlan:2023}, influencing their host galaxy evolution. This is in addition to being useful probes of star formation under extreme conditions at galactic centres \citep{Morris:1996, Barnes:2017}.

Current studies on the formation and growth of NSCs agree on mostly two dominant pathways for NSC growth. NSCs can grow either via in-situ star formation in the central region of galaxies, fuelled by the accretion of gas onto the central region \citep{Milosavljevic:2004, Seth:2008, Antonini:2015, Kacharov:2018}; or they can grow due to the infall of globular clusters (GC) that sink to the central region due to dynamical friction and end up fully disrupted, contributing to the total NSC mass \citep{Tremaine:1975, Capuzzo-Dolcetta:1993, Capuzzo-Dolcetta:2008, Gnedin:2014, Tsatsi:2017}. 

These different formation channels are expected to leave distinct signatures in the NSC properties, from their ages and metallicities to the internal kinematics. For instance, the GC infall naturally explains the presence of a population of metal-poor stars found within NSCs \citep{Arca-Sedda:2020,Feldmeier-Krause:2020, Fahrion:2020, Pinna2026}. On the other hand, an in-situ formation scenario would result in an NSC hosting a population with a more complex chemistry and star formation history, and potentially with a prominent younger population if there is recent star formation \citep{Kacharov:2018, Fahrion:2019, Hannah:2021}. Moreover, stellar kinematics provides additional information, as the two pathways are expected to imprint distinct signatures on the NSCs. For instance, slowly or non-rotating NSCs are mostly associated with the GC infall scenario, while rotating NSCs are typically associated to in-situ formation since the gas funnelled to the central region carries angular momentum. However, the GC infall scenario cannot be ruled out in the presence of rotation. Additionally, these signatures can also depend on the host galaxy type \citep{Seth:2010, Lyubenova:2019, Pinna:2021}. However, resolved kinematic measurements of NSCs are difficult to obtain, particularly in the dwarf galaxy regime \citep{Neumayer:2020}. However, recent studies have made use of integral-field spectroscopy to infer the stellar kinematics of unresolved NSCs, including in the dwarf galaxy regime, successfully determining the dominant formation channels in various galaxies \citep{Fahrion:2020, Fahrion:2022, Johnston:2020}.

Nevertheless, to be able to explain all the features presented across various NSCs, a combination of these two scenarios is often invoked \citep{Lyubenova:2013, Antonini:2015, Guillard:2016}. In this context, recent works discuss a correlation between the host galaxy mass and the dominant growth channel \citep{Neumayer:2020, Fahrion:2021, Fahrion:2022}, suggesting a transition from mainly in-situ to accretion at $M_\star \approx 10^9 \msun$. Then, NSCs in systems below this mass scale tend to be mostly dominated by old stellar populations, hence supporting the idea of the accretion of GCs dominating the NSC growth in dwarf galaxies, in particular. This mass dependence is also reflected in the fraction of galaxies hosting an NSC (i.e. nucleation fraction, $f_n$), which peaks at $M_\star \approx 10^9 \msun$, declining significantly at lower masses \citep{Sanchez-Janssen:2019, Hoyer:2021}. This would be consistent with the transition in the dominant growth channel, as well as the reduced GC abundance that drives the NSC growth in these systems.

While the main growth channels for NSCs are relatively well understood, their early formation still remains uncertain \citep{Neumayer:2020}. This is both due to the difficulty in obtaining reliable observational measurements to address this and, in the case of simulations, due to the level of complexity in the models and the computational cost of running them in a cosmological setting. Nevertheless, in the recent years, various works address the formation of NSCs in isolated settings \citep[e.g.][]{Guillard:2016, Jo:2026} or in a limited cosmological setting \citep{Garcia:2025}, without reaching to $z=0$. Recently, using the EDGE simulations, \citet{Gray:2025} explored the formation of NSCs finding that their formation can be triggered during galaxy-galaxy mergers in the dwarf galaxy regime, further expanding our understanding of NSC formation in a full cosmological context. 

In this work, we use the LYRA cosmological simulations \citep{Gutcke:2022-LYRAII, Gutcke:2022-LYRAIII}, with a sample of six dwarf galaxies with $M_\star=4\times10^5-10^7 \msun$ ($\mathrm{M}_{200\mathrm{c}} = 7\times 10^8 - 5\times 10^9 \,\msun$\footnote{$\mathrm{M}_{200\mathrm{c}}$ corresponds to the DM mass enclosed within $R_{200\rm c}$, i.e. the radius where the mean enclosed density is 200 times the critical density of the Universe.}) to study the formation and growth of NSCs in dwarf galaxies. This sample of dwarfs contains galaxies totally quenched by reionisation, as well as star forming dwarfs. The latter can significantly grow in stellar mass in the central region, suggesting an important role of in-situ star formation in the growth of NSCs in these dwarfs. In addition, we highlight the role of mergers in the mass growth of the quenched dwarfs \citep[see][]{Sureda:2026}, which can potentially contribute to the formation of NSCs. 

This paper is organised as follows. Section~\ref{sec:methods} briefly describes the LYRA model and the methods used to identify the NSCs and to associate an origin for each star particle within the NSCs. Section~\ref{sec:results} presents the main results of this work regarding the NSC properties, their emergence, and their main growth channels, and includes some properties of the stellar populations associated with the different channels. In Section~\ref{sec:discussion} we further discuss some of the implications of our results and place them in the context of NSC observations. Finally, Section~\ref{sec:conclusions} presents the main conclusions of this work.

\section{Methods}\label{sec:methods}

\subsection{LYRA model}

The simulations used in this study are based on the LYRA galaxy formation model \citep[for full details of the LYRA model refer to][]{Gutcke:2021-LYRAI, Gutcke:2022-LYRAII}, implemented in the cosmological hydrodynamical moving mesh code \texttt{AREPO} \citep{Springel:2001, Pakmor:2016, Weinberg:2020}. The LYRA model is able to follow the formation and evolution of individual stars down to $4\,\msun$ thanks to the inclusion of a resolved multiphase ISM with cooling prescriptions down to $10$ K. This in turn allows for completely resolved supernovae blastwaves. To model the metal enrichment at high redshift, a Pop III star metal enrichment is included as part of a subgrid model \citep{Gutcke:2022-LYRAII}. In practice, this incorporates a metallicity enhancement (from 0 to $10^{-4} \, \mathrm{Z_\odot}$) in the gas within the virial radius once a halo the reaches $10^6 \msun$, and its then allowed to cool and subsequently form stars. The ultra-violet background (UVB) is implemented as an homogeneous and isotropic heating term following \citet{Faucher:2020}, setting the HI reionisation to start at $z=7.8$ and complete by a $>90\%$ by $z=7.3$. The star formation prescription follows a Schmidt relation \citep{Schmidt:1959}, with efficiency increasing as a function of the gas density, reaching a $100\%$ efficiency at $n_\mathrm{H}>10^4 \mathrm{cm}^{-3}$. Finally, the stellar masses are individually sampled from a Kroupa IMF \citep{Kroupa:2001} down to $M_{*,\min}=4\,\msun$. Below this threshold, star particles represent the integrated population. 

The zoom-in initial conditions are drawn from the 100 Mpc DMO box of the EAGLE project \citep{Schaye:2015}. The cosmology parameters used for the runs are consistent with Planck 2013 results \citep{Planck:2014}, i.e. a flat universe with $h=0.6777$, $\Omega_{\rm m}=0.307$, $\Omega_{\rm bar}=0.048$, $\Omega_{\rm \Lambda}=0.693$, $\sigma_8=0.8288$. The DM mass resolution within the high-resolution region is $~80 \msun$. The target gas mass resolution in this region is set to $4\,\msun$ and is enforced by the \texttt{AREPO} refinement and de-refinement routines. We construct the merger trees based on both the stellar and DM component using the \texttt{D-halos} tree code \citep{Jiang:2014} to be able to follow more accurately the growth histories of the simulated haloes. This allows for the proper identification of significant mergers that might have an impact on the overall properties of these galaxies, and potentially contribute to the NSC evolution.

\subsection{Nuclear Star Cluster identification}\label{sec:NSC-identification}

Nuclear star clusters in the simulation are identified via fitting the stellar surface density profile of the galaxies. In observations, this fit would be done to the surface brightness profile, however the mass profile is a good tracer of the light profile when there is no ongoing star formation, and we choose not to model the light distribution of the galaxies in order to keep the intrinsic stellar distribution. Each profile is fit with one and two Sérsic components,

\begin{equation}
    \Sigma_\star^{\mathrm{single}}(r;\theta) = A \exp{\left\{-b_n\left[\left(\frac{r}{r_e}\right)^{(1/n)} - 1\right]\right\} },
\end{equation}
where $\theta = (A, r_e, n)$ is the set of free parameters of the model, $A$ is the amplitude, $r_e$ is the effective radius and $n$ is the Sérsic Index. The constant $b_n$ is defined such that $r_e$ contains half the total luminosity. Then we have two different models to fit of the data, the single-Sérsic and the double-Sérsic. To quantify which model is preferred, we compute the Akaike Information Criterion \citep[AIC;][]{Akaike:1974}, as $\mathrm{AIC} = 2k - 2\ln{(\hat{\mathcal{L}})}$, where $k$ is the number of free parameters and $\hat{\mathcal{L}}$ is the maximum log-likelihood of the model. In this case, we are fitting the data using a least-square-sum regression, therefore the log-likelihood is determined by $\ln{(\hat{\mathcal{L}})} = -\frac{n}{2}\ln{\frac{\mathrm{RSS}}{N}} + C$, and the AIC statistic becomes 

\begin{equation}
    \mathrm{AIC} = 2k + N\ln\left({\frac{\mathrm{RSS}}{N}} \right) + C,
    \label{eq:AIC}
\end{equation}
where $\mathrm{RSS}$ is the sum of squared residuals, $N$ is the sample size and $C$ is a constant independent of the model that later gets cancelled out. In this way, the difference between the AIC values gives a measure of the relative fitting quality between the two models, favouring the model with the lowest AIC, which is then selected as the preferred model. The advantage of using an information criterion as the AIC statistic is that we penalise models with larger number of parameters via the factor $2k$, potentially preventing the preference of a model that overfits the data. Finally, we establish the presence of an NSC in a galaxy if the double-Sérsic model is preferred. We perform this routine for each galaxy at every snapshot so we can identify the emergence of the NSC. In addition, we use the fit information to characterise the NSC region and determine the NSC radius at $z=0$, denoted as $r_\mathrm{NSC}$, as the intersection of the two Sérsic components where the inner component becomes insignificant in comparison to the outer component as the radius increases. Throughout this work, $r_\mathrm{NSC}$ refers to the measurement at $z=0$.

\subsection{Star particle labelling}\label{sec:particle-label}

The main objective of this work is to determine the origin of the stars that make up the identified NSCs. To achieve this, we select the particles within the NSC radius at $z=0$ and classify them according to their origin. In particular, we use the position of the star particles and their subhalo membership at the first snapshot after birth to classify them accordingly. Throughout this work, we will use the following categories to label the origin of the different star particles.

\subsubsection{NSC-Born}
Stars in this category form in situ through gas accretion to the centre of the galaxy. To determine this, we inspect the position of the star particle at the first snapshot after birth and compare it to the centre of the main galaxy. If the distance to the centre is smaller than the NSC radius, we then label these star particles as \textit{NSC-Born}. We stress that the stars in this category are defined based on a hard radial cut. In reality, the Sérsic profiles of the galaxy and the NSC overlap, which means there are overlapping populations of stars from both components near this boundary. We address this by including another category that mostly captures stars that could otherwise be missed by this radial cut.

\begin{figure}
    \centering
    \includegraphics[width=0.99\linewidth]{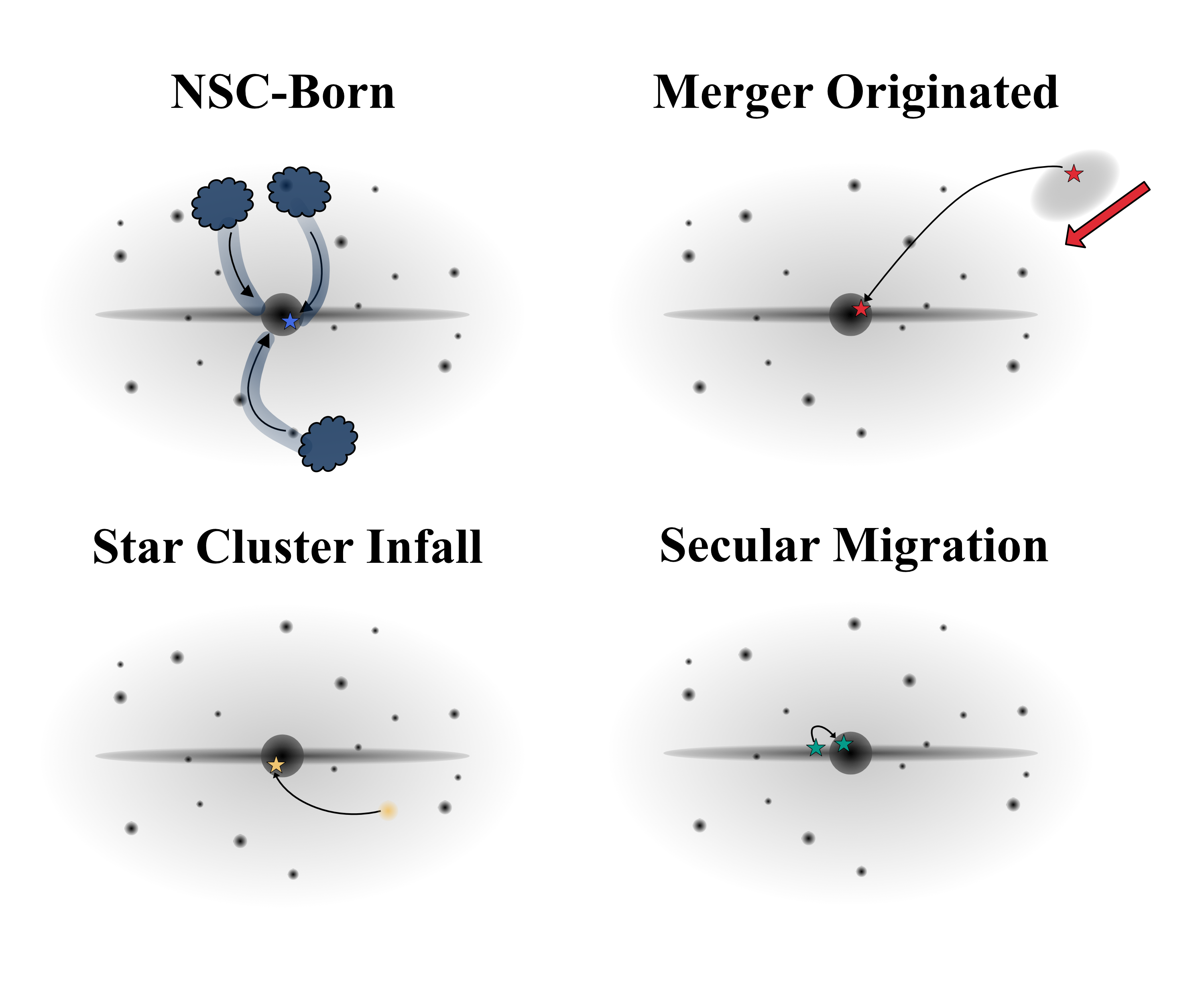}
    \caption{Descriptive illustration of the different growth channels investigated in this work. \textit{Top-Left:} NSC-Born stars, determined by their position after birth. \textit{Top-Right:} Merger contributed stars born in a satellite galaxy. \textit{Bottom-Left:} Star cluster infall that disrupt and contribute some of their mass to the NSC. \textit{Bottom-Right:} Secular migration of stars in the vicinity of the NSC.}
    \label{fig: Illustration}
\end{figure}

\subsubsection{Merger originated}
We classify as \textit{Merger Originated} the stars born in a different system that was accreted into the main halo and end up in the NSC population at $z=0$. For this, we look at all the progenitors that share particles with the NSC. Then we go back to each progenitor branch and compute the contribution to the NSC mass at the moment when they have the largest $M_\star$ across their whole branch. Since the selection relies on having at least one shared particle with the NSC, we also include in this category any star with a birth position outside $R_{200c}$ of the main halo at the time of birth. This is a minor correction that constitutes up to $\sim 5\%$ of the mass budget in this category. In this way, we are able to consistently associate more than $95\%$ of this mass budget to individual progenitors, while the remaining mass is associated to accreted field stars.

\subsubsection{Star cluster originated}

Since we are interested in studying the impact of star clusters on the formation of the NSCs, we need to define a method to identify said star clusters. In this work, we will make use of the \texttt{Subfind} structure finder to identify dark matter deficient substructures. We consider a subhalo to be a star cluster candidate if $M_\mathrm{DM}/M_\mathrm{tot} < 0.5$. Additionally, we want to ensure we are not capturing smaller systems stripped from their dark matter halo \citep[see][]{Gutcke:2024}, hence being only temporarily stellar-dominated systems. To this end, we use the merger trees to inspect the full history of the star cluster candidates and only select candidates that match the above criterion throughout their entire lifetimes, i.e. structures that have never been dark matter dominated. We stress that this is especially relevant to distinguish between the contribution from the accretion of smaller galaxies and the accretion of star clusters onto the NSC. In this work, we have not modified \texttt{subfind} to improve the identification of bound star clusters. However, we ran an additional set of merger trees that follow the stellar components, allowing us to identify and follow the history of these DM-deficient subhaloes, including those entirely free of DM, throughout their entire branch. Finally, we classify as \textit{star cluster originated} the stars associated with these structures before merging with the NSC. Furthermore, if a satellite galaxy enters the main halo with one or more star clusters associated to it, and these star clusters are subsequently accreted onto the NSC of the main halo, then those stars are labelled as \textit{star cluster originated}.

\begin{figure}
    \centering
    \includegraphics[width=0.99\linewidth]{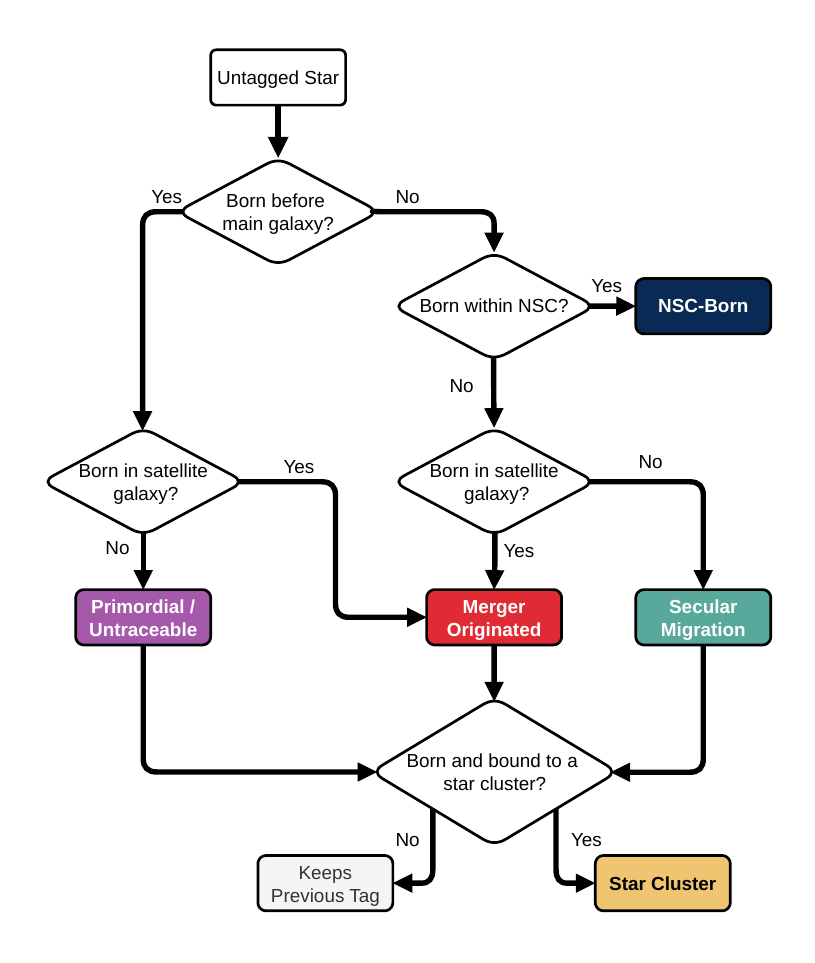}
    \caption{Particle labelling workflow diagram. For each star particle we start with an unlabelled particle and follow the decision tree until we reach a label.}
    \label{fig : tagging diagram}
\end{figure}

\subsubsection{Secular migration}
Some NSC stars form within the host galaxy but not originally in the NSC region. Then, during the evolution of the galaxy, these stars can migrate to the central region and be identified as part of the NSC at $z=0$. In many cases, this population is composed of stars that form close to the NSC radius and end up orbiting the NSC boundary. Hence, these stars represent an in-between population that can be easily excluded with a different radial definition of the NSC, as noted in the definition of the \textit{in-situ} component. Finally, note that NSC stars formed during merger interactions also fall under this category if they are first associated with the main progenitor by \texttt{subfind}.

\subsubsection{Untraceable stars}
It can happen that some NSC stars are formed before we can properly identify the main branch of the galaxy, usually at $z>10$. If we are not able to associate these stars with any progenitor of the main halo or any star cluster, then we label them as untraceable. While these comprise a negligible fraction of the NSC mass ($< 10$ particles) in all cases, we still preserve this label on the star particle list to exclude them from any other category. However, we do not consider them for most of the analysis presented in this work.

\subsubsection{Workflow for the particle labelling}

With our categories defined, we start by inspecting each individual star in the NSC. Based on the star particles age and birth position, we determine if the stars were born within the NSC, or within the host galaxy or if they were born in an external system. We categorise them as \textit{NSC-Born}, \textit{Secular Migration} or \textit{merger originated}, respectively. Next, we take all stars that are not labelled as \textit{NSC-Born} and check whether they were born in identified star clusters. If that is the case we update their label to \textit{star cluster originated}. This last step allows for identifying star clusters born either within the main halo or in a satellite system. The full procedure is presented in Fig.~\ref{fig : tagging diagram} as a flow diagram for clarity.

\section{Results}\label{sec:results}

We present our results from the analysis of the stellar components of the LYRA galaxies, focussing first on the NSC properties, the emergence of the NSCs, and the origin of the stars within the present-day NSC. Then, we analyse the properties of the stellar populations originating from different sources and discuss potential ways to distinguish them observationally. 

\subsection{Nuclear star cluster properties}\label{sec:NSC-properties}

\begin{figure*}
	\centering
    \includegraphics[width=0.99\linewidth]{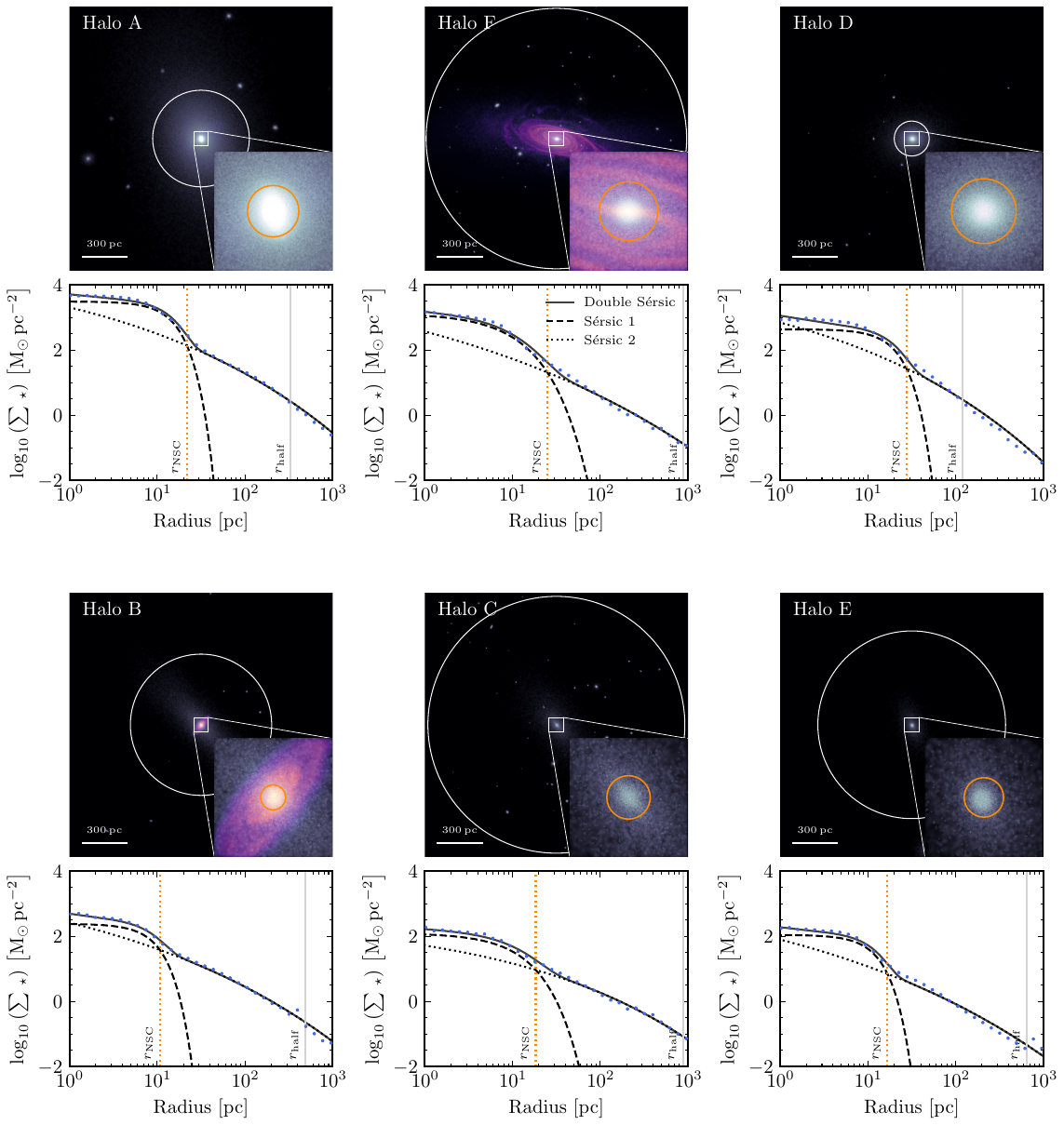}
    \caption{Surface density images of the gas (purple) and star particles (white) of the LYRA galaxy sample at $z=0$. The inset shows a zoom to a $100\times100$ pc region at the centre.  Each bottom panel shows the surface density profile with the corresponding double Sérsic fit. The NSC radius (orange) and the stellar half mass radius (grey) are shown as circles in the top panels and as vertical lines in the profile.}
    \label{fig: NSC surface density profile}
\end{figure*}

Figure~\ref{fig: NSC surface density profile} shows the $z=0$ surface density images of the galaxies in the sample. This illustrates the general structure of the dwarfs, highlighting the presence of NSCs shown in the insets for each galaxy. At the bottom of each image, the radial surface density profile of each galaxy is shown, featuring the characteristic increase of the surface density towards the inner region of the galaxy due to the presence of an NSC. We note that, at this stage, we have not aligned the galaxies in any particular way. This could lead to the profiles being biased if the NSCs are not spherical. This might be the case for Halos F and B, both of which show a clear gas disc, since the stellar component could inherit its structure from the gas dynamics. However, we verified that the stellar component is indeed spherical within the inner $\sim 50$ pc for all haloes except Halo C. This halo has a more elongated shape, as we can see from the image; however, the mass excess towards the centre remains clearly visible despite this asymmetry.

We then fit the profiles, finding the double-Sérsic model is preferred in all cases based on the AIC criterion (see Sec.~\ref{sec:NSC-identification}). The double-Sérsic fit to the profile is shown (solid line) along with the individual components (dashed and dotted lines). The nuclear star cluster radius, $r_\mathrm{NSC}$, is then identified as the radius where both components intersect and the inner Sérsic component becomes dominant towards the inner region. We note that due to how we define the NSC radius, the values listed here are slightly large in comparison to observed NSCs. We tested using the effective radius of the central inner Sérsic profile or the NSC half mass radius, and these generally result in somewhat smaller radii. Indeed, regardless of how the radius is defined, we can confirm that the substantive results remain consistent.

With the NSC radius defined, we can use these values to extract more information from the present day NSCs. In Table~\ref{tab:NSC parameters}, we show both the host galaxy and the NSC mass, the NSC radius, the mean mass-weighted metallicity and age of each NSC. We also show the fraction of old metal-poor stars, $f_\mathrm{OMP}$, defined here as the fraction of stars with $\mathrm{Age} > 2 \,\mathrm{Gyr}$ and $\log_{10}\left(Z/Z_\odot\right) < -1.0 \,\mathrm{dex}$ as used in other works \citep{Fahrion:2022}. The last column shows the mass fraction of stars born within the NSC, $f_\mathrm{NSC-Born}$, defined using the full particle data. Interestingly, the last two columns of our table already highlight an important point: our data does not support the notion that old metal poor stars alone are a good tracer of the fraction of stars accreted from GCs in all instances. If this were the case, we would expect $f_\mathrm{OMP}$ to be small when $f_\mathrm{NSC-Born}$ is high. However, in the case of Halo E, it has the largest $f_\mathrm{NSC-Born}$ but does not have a particularly small $f_\mathrm{OMP}$. Indeed, it is in the regime of being classified as dominated by GC accretion.

\begin{table*}
    \begin{tabular}{lccccccc}
    \hline
    Name & $\log_{10}\left(M_\star/\mathrm{M}_\odot\right)$ & $\log_{10}\left(M_\mathrm{NSC}/\mathrm{M}_\odot\right)$ & $r_\mathrm{NSC} \, \left[\mathrm{pc}\right]$ & $\left<Z/Z_\odot\right>$ & $\left<\mathrm{Age}\right> \, \left[\mathrm{Gyr}\right]$ & $f_\mathrm{OMP}$ & $f_\mathrm{NSC-Born}$ \\
    \hline
    Halo A & 6.98 & 6.12 & 21.85 & -0.63 & 11.78 & 0.32 & 0.89 \\
    Halo F & 6.39 & 5.44 & 24.91 & -0.98 & 8.56 & 0.65 & 0.82 \\
    Halo D & 6.20 & 5.56 & 27.48 & -0.77 & 12.78 & 0.58 & 0.78 \\
    Halo B & 6.10 & 4.67 & 10.71 & -1.31 & 11.85 & 0.80 & 0.32 \\
    Halo C & 6.02 & 4.51 & 18.40 & -0.86 & 13.56 & 0.76 & 0.75 \\
    Halo E & 5.60 & 4.53 & 16.72 & -0.67 & 13.54 & 0.68 & 0.95 \\
    \hline
    \end{tabular}
    \caption{Present day parameters of the galaxies and their nuclear star clusters. The columns show the name of the halo, the galaxy stellar mass, the NSC mass, the NSC radius, mass weighted metallicity and age, the fraction of OMP stars and of stars born within the NSC. The haloes in the table are sorted by decreasing stellar mass.}
    \label{tab:NSC parameters}
\end{table*}

\begin{figure*}
    \centering
    \includegraphics[width=0.8\linewidth]{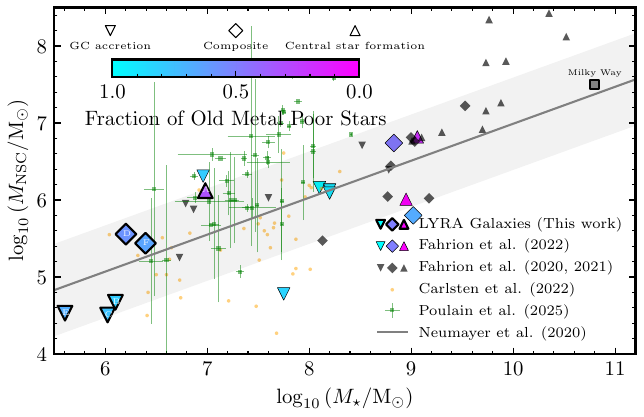}
    \caption{NSC-galaxy mass relation for the LYRA dwarfs, encoding information about the fraction of old metal poor stars in the NSCs. For a direct comparison we show data of observed NSCs from \citep{Fahrion:2020, Fahrion:2021, Fahrion:2022}, with their dominant growth channel indicated by the marker shape as indicated in the labels. Other points correspond to recent observations of NSCs in dwarf galaxies from the ELVES \citep{Carlsten:2022-NSCs} and MATLAS \citep{Poulain:2025-MATLAS} surveys. The line and shaded region correspond to a relation shown in \citet{Neumayer:2020}. We see that LYRA NSC masses are within the expectations for this mass regime, with a dominant population of old metal-poor stars in most galaxies. }
    \label{fig:NSC mass vs galaxy mass with literature}
\end{figure*}

Additionally, we inspect some scaling relations between the NSC and the host galaxy. In particular, we show the galaxy-to-NSC mass relation in Fig.~\ref{fig:NSC mass vs galaxy mass with literature}, in comparison with different predictions and observational samples from the literature. The relation is derived from samples of spectroscopically and dynamically modelled NSC masses, as well as masses derived from stellar population models, as presented in \citet{Neumayer:2020} and references therein. We also show recent samples of unresolved NSCs in dwarfs from the ELVES \citep{Carlsten:2022-NSCs} and the MATLAS \citep{Poulain:2025-MATLAS} surveys. Additionally, we include the fraction of old metal poor stars, estimated via spectral fitting analysis from \citet{Fahrion:2020, Fahrion:2021, Fahrion:2022}, which also include an inferred dominant growth channel, allowing us to directly compare our derived $f_\mathrm{OMP}$ values. 

We find that the NSCs found in the LYRA sample follow the expected trend for this relation in the dwarf galaxy regime. Moreover, the obtained fraction of old metal poor stars is slightly higher than expected in some cases, with the majority of the NSCs dominated by an old metal poor population (i.e. high $f_\mathrm{OMP}$ values), as found for this mass regime. However, for the dwarfs able to sustain star formation after reionization \citep[Haloes A, F and D; see][]{Gutcke:2022-LYRAIII, Sureda:2026}, this fraction is significantly smaller than for the rest of the sample, indicating that at least some of this star formation occurs in the central region of the galaxy.

\begin{figure*}
    \centering
    \includegraphics[width=0.75\linewidth]{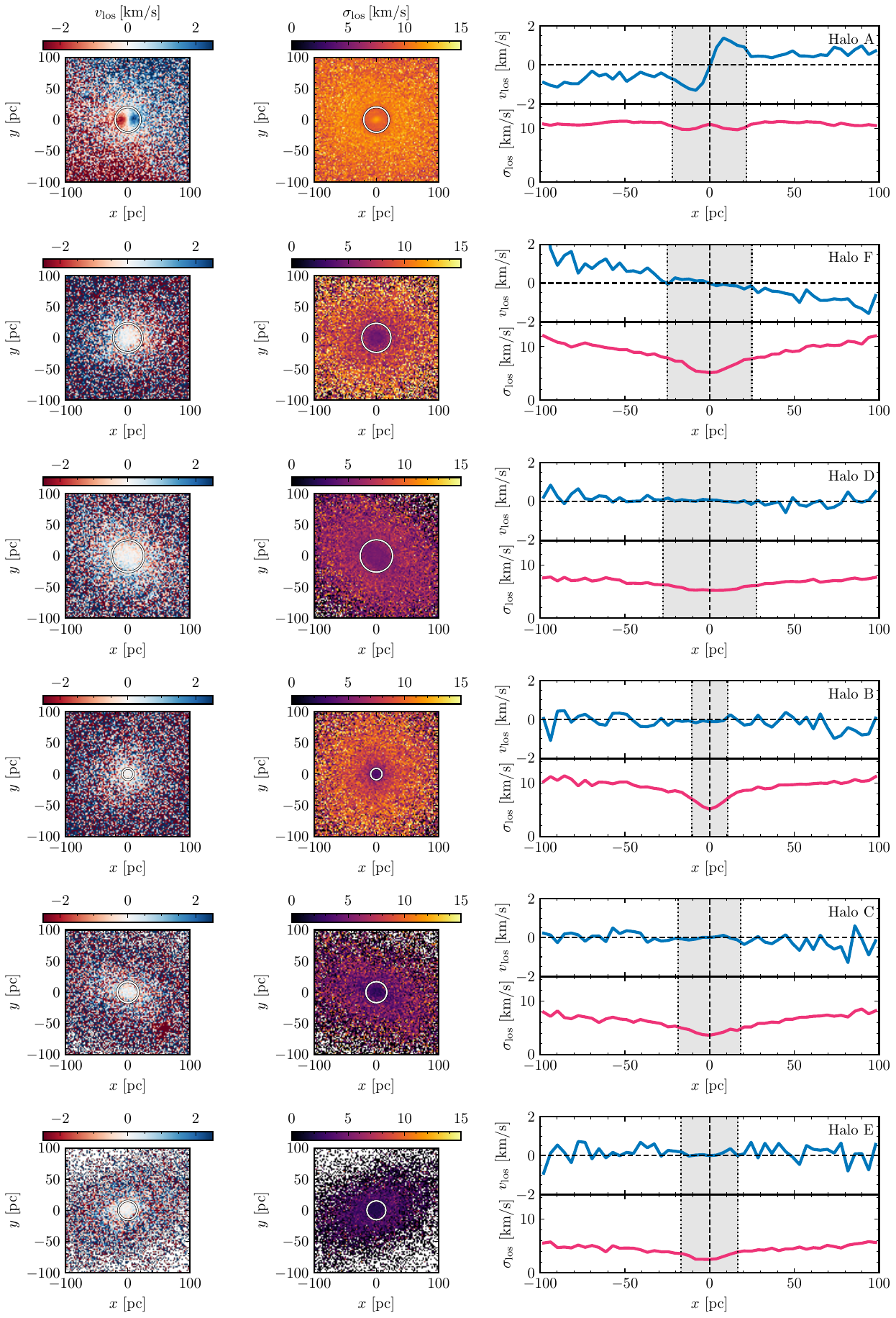}
    \caption{Line-of-sight velocity (left) and velocity dispersion (middle) maps in a $100 \times 100$ region, and kinematics profile along the x-axis (right) for each halo. The NSC region is indicated by a circle in the left and middle panels, and as a shaded band in the right panel. The haloes are rotated such that the y-axis is aligned with the direction of the angular momentum vector of the stars bound to the main halo, i.e. an edge-on projection. All the haloes within this region and their NSCs appear as dispersion-dominated. Only Halo A and Halo F exhibit signs of ordered rotation, with only Halo A showing it within the NSC region.}
    \label{fig:vdisp}
\end{figure*}

The stellar kinematics of the central region of a galaxy contains valuable information that can be linked to the formation of its NSC. While in non-dwarf, larger-mass systems we expect the overall rotation to be consistent with the host galaxy, smaller and spheroidal systems have a more complex kinematic structure that allows for less rotation or even counter-rotation with respect to the host galaxy \citep{Seth:2008,Schodel:2009,Lyubenova:2013,Neumayer:2020}. In Fig.~\ref{fig:vdisp}, we show the line of sight velocity, $v_\mathrm{los}$, and velocity dispersion, $\sigma_\mathrm{los}$, of a $100 \times 100$ region centred on the NSC (shown as the white circle) for all haloes. The haloes are rotated such that the angular momentum vector of the star particles bound to the main halo is aligned with the y-axis. On the right we also show the radial profiles along the x-axis of both $v_\mathrm{los}$ and $\sigma_\mathrm{los}$. To compute this radial profile, we select particles within $\pm50$ pc on the y-axis to focus on the NSC component (shown as the shaded region). We have additionally verified that these are indeed self-gravitating by inspecting the their circular velocity profiles, which indicate that the stars dominate the mass budget in this region. 

From the Fig.~\ref{fig:vdisp}, we see that these are all dispersion-dominated systems with a typically small or even negligible rotational velocity component. We highlight that only haloes A and F display signs of rotational velocity, although it is small in magnitude compared to the velocity dispersion. Moreover, only Halo A shows this rotation signature within the NSC region. Interestingly, for this halo, both the host galaxy and the NSC have a non-negligible rotational component as we can see in its $v_\mathrm{los}$ map. However, the rotation axis of the NSC region and the host are not completely aligned as the rotation axis of the NSC seems to be tilted with respect to the one of the host, highlighting the complex kinematics of this structure. While we do not perform any further analysis on the source of this rotational signature, we want to point out that this is the most massive galaxy in our sample, only followed by Halo F. This at least hints at the fact that the presence of a rotating component may be a feature favoured at larger stellar masses. 

All haloes in our sample, except for Halo A, exhibit a drop in the velocity dispersion close to the NSC region. Such a feature has been discussed in the context of galactic bars, nuclear stellar rings, as well as NSCs \citep{Comeron:2008_sigma_drops, Portaluri:2017, Lyubenova:2019}. In the case of NSCs, a $\sigma$-drop without evidence for further ordered rotation is seen as a feature of a compact, bound component which is embedded in a kinematically hot component, in this case, the NSC within the rest of the galaxy. This feature is usually associated with the growth of an NSC via the accretion of GCs, which would leave a self-gravitating component in the centre, kinematically distinct from the rest of the galaxy \citep{Hartmann:2011, De-Lorenzi:2013, Antonini:2013}. If, however, the NSC grows mostly via in-situ star formation, it is expected that the stellar orbits within the NSC inherit their angular momentum from the infalling gas that fuels star formation. This would then appear as a rotating stellar component in the central region of the galaxy, further reducing the velocity dispersion in the region \citep{Seth:2008,Guillard:2016}. While this $\sigma$-drop can be long-lived, it requires further gas accretion and subsequent star formation at a rate of $1 \msun \,\mathrm{yr}^{-1}$ in order to survive \citep{Wozniak:2006_sigma_drops}. This can easily occur in larger mass systems featuring prominent discs, galactic bars and spiral arms. However, for low-mass systems it is more difficult to fuel gas to the central region to support this continuous level of star formation. 

In our sample, we can conclude that, in most cases, the majority of the mass within the NSC is formed inside that same region, i.e., in-situ dominated NSCs. Therefore, the stellar kinematics suggest that this in-situ component was in place at early times and any signature of rotation has been erased. The fact that the NSC in Halo A does display mild rotation signatures may be related to the more recent in-situ star formation in this halo, as we discuss in Sec.~\ref{sec:mass_dist}  \citep[see also,][]{Sureda:2026}. This is also supported by the lower fraction of old metal-poor stars present in this halo. We then want to explore how the NSCs in these dwarfs assemble and grow in a way that leaves these signatures in both the $f_\mathrm{OMP}$ and the stellar kinematics.

\subsection{Emergence of nuclear star clusters in LYRA}\label{sec: emergence}

To investigate how these NSC emerge, for each galaxy we inspect their stellar surface density profiles at different times. We then use the AIC criterion from Eq.~\ref{eq:AIC} to quantify if the profile requires a single or a double Sérsic component to be modelled, and by doing this over time, we estimate the moment when the nuclear region becomes prominent. Since the transition from non-nucleated to nucleated is not instantaneous, it is not possible to identify a specific time for the emergence of an NSC. However, we identify a period of time where the transition occurs and there is a clear difference in the profiles before and after this emergence window. The width of the window is selected case by case, based on how consistent the preference for a single component model vs a double component model is, given by the AIC statistic. Across our sample this emergence window has average redshifts in the range $z\sim 12.9 - 5.6$ ($0.34-1.05\, \mathrm{Gyr}$), implying that these NSCs are in place at early times.

\begin{figure}
    \centering
    \includegraphics[width=0.99\linewidth]{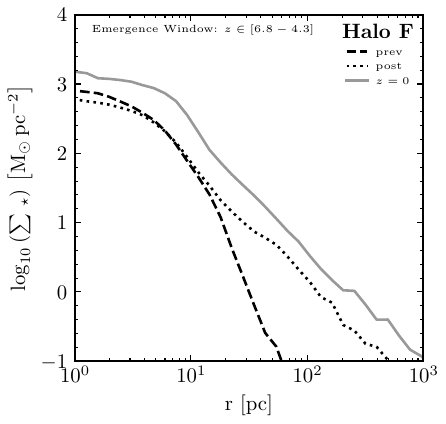}
    \caption{Stellar surface density profile at different times for Halo F. In dotted (dashed) lines we show the median of the profiles after (prior to) the NSC emergence window, around $z \sim 5.6$. In a light solid colour we also show the profile at $z=0$ for comparison. This highlights the emergence of the nucleated feature in the profile, starting already with the compact component and then growing in the outskirts as time passes.}
    \label{fig:emergence}
\end{figure}

To understand how the NSC emerges during the galaxy assembly history, we compute the median of the surface density profiles for snapshots before and after this emergence window. We show this for Halo F in Fig.~\ref{fig:emergence}. The dashed and dotted lines show the median profile before and after the emergence window, respectively. The previous and subsequent intervals are selected to contain the same number of snapshots to ensure a comparably smooth median estimate in both intervals. We note that this implies that the medians are computed over different physical time spans for the two bins; however, this choice ensures that both medians are equally reliable for comparison, better highlighting the separation between the pre-NSC and post-NSC surface density profiles. For comparison, the $z=0$ profile is shown as a solid line. From this figure we can infer that the NSC feature in the surface brightness profile emerges as the galaxy grows towards the outskirts. More importantly, this shows that the main NSC structure is already seeded early on and with time the rest of the galaxy grows around and together with the compact system. Such a scenario has already been discussed in the context of the initial formation of NSCs \citep[see Section 7.1 in][]{Neumayer:2020}.

Most haloes in LYRA show similar profiles as the one shown for Halo F (we show a similar plot for all haloes in Appendix~\ref{app:emergence_all}), highlighting a common pattern for the formation of NSCs. The exception in our sample is Halo B, which does show an increase in the surface density of the inner region in addition to the extension at larger radii. This might hint towards a distinct formation path for this Halo and we will expand this discussion in Sec.~\ref{sec:OriginOfNSCStars}. Nevertheless, other works point towards other formation mechanisms in play. For instance, \citet{Garcia:2025} discusses how star clusters formed from infalling gas can merge together and give rise to a NSC. In addition, using the EDGE simulations, \citet{Gray:2025} discusses the formation of NSCs through starbursts driven by galaxy mergers in the dwarf galaxy regime. This apparent diversity of NSC formation mechanisms highlights the complexity of this process and the need for exploring in more detail the resulting stellar populations found in the present-day NSCs.

\subsection{Origin of NSC stars}\label{sec:OriginOfNSCStars}

\begin{figure}
    \centering
    \includegraphics[width=0.95\linewidth]{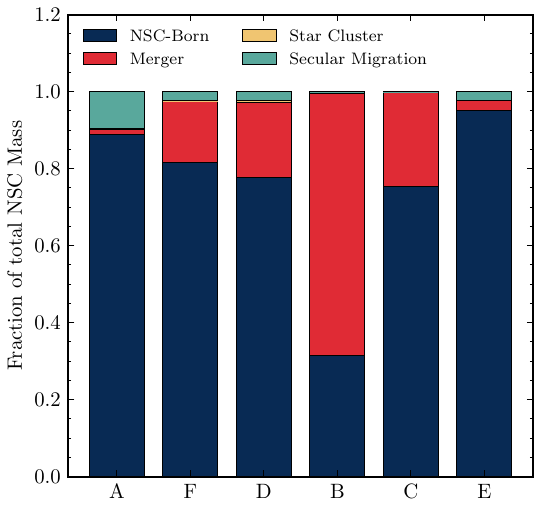}
    \caption{Fraction of NSC mass from different formation channels for all LYRA haloes. Most haloes are dominated by the \textit{NSC-Born} component (i.e. in-situ stars), followed by the \textit{Merger} contribution. In all haloes, the \textit{Star Cluster} contribution is negligible.}
    \label{fig:Fractions}
\end{figure}

One of the key objectives of this work is to understand the main growth channels of NSCs. Therefore, we focus on classifying the $z=0$ star particles of the NSCs according to their origin. With the particle labelling methodology described in Section~\ref{sec:particle-label}, we associated each star particle to a different formation channel. We then obtain the fractional mass contribution of each of these populations, as shown in Fig.~\ref{fig:Fractions}. We immediately see that the all of the NSCs in our sample, except Halo B, are strongly dominated by stars born within them, i.e., the \textit{NSC-Born} component, with its value already listed in Table~\ref{tab:NSC parameters}. While this seems at odds with our current understanding of NSC formation in dwarf galaxies, it is important to mention that our definition of \textit{NSC-Born} star is deterministic and is based only on the position of the corresponding star particle at birth. In contrast, observations define their in-situ population based on inferred stellar properties, i.e. stellar kinematics, age and metallicity estimates from the observed spectra, implicitly making the assumption that in-situ star formation occurred more recently. In fact, when we split the stars labelled in the \textit{NSC-Born} component by their stellar ages, we find that only a small fraction of this component is made of recent in-situ star formation and would otherwise be classified as an old stellar population. This aligns more closely with the evidence found in observations and explains why the LYRA NSCs can have a large fraction of old metal-poor stars, as well as high \textit{NSC-Born} mass fractions (see Fig.~\ref{fig:NSC mass vs galaxy mass with literature}). Throughout this work, we will refer to the \textit{in-situ} (or \textit{NSC-Born}) population as the one derived from our particle data, as described in Section~\ref{sec:particle-label}.

\begin{figure}
    \centering
    \includegraphics[width=0.98\linewidth]{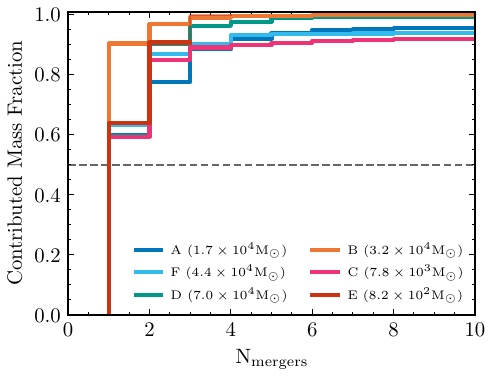}
    \caption{Contributed mass from different progenitors into the \textit{merger} component of each NSC. The progenitors are sorted by decreasing mass contribution. The values shown in the labels are the NSC mass in the \textit{merger} category.}
    \label{fig:mergers}
\end{figure}

The second most relevant component of the NSCs in our sample is usually the \textit{Merger} one. This is especially relevant in galaxies with little to no star formation after reionization, since these galaxies grow only in stellar mass through the mergers of smaller systems \citep[see][]{Sureda:2026}. In particular, for Halo~B, this is the dominant contribution, reaching $>60\%$ of the NSC mass. We further inspect the contribution from individual progenitors into the merger contributed mass of each NSC. For this, in Fig.~\ref{fig:mergers} we show the cumulative contribution from individual mergers, to the NSC mass in this channel (the actual value is shown in the labels), sorted by decreasing contributed mass. We note that most of the contribution from this channel is produced by the merger of one satellite with the main halo, in each case contributing more than $50\%$ of the merger mass budget. We note that Halo~B shows a substantially larger mass contribution from a single merger, reaching $\sim90\%$ of the merger mass from this single progenitor, implying that the majority of the NSC mass in this halo is contributed by this single merger.  

We next consider the \textit{secular migration} channel. Its contribution is generally small, accounting for less than $10\%$ of the NSC mass and, in most cases, less than $5\%$. However, most stars assigned to this category form in close proximity to the NSC, making their classification sensitive to the adopted NSC radius. Our use of a sharp radial cut is primarily a practical choice and does not capture the intrinsically overlapping nature of the NSC and its surrounding galaxy. Consequently, the values reported here likely overestimate the contribution from secular migration, although this channel is clearly non-zero. Nevertheless, the presence of this distinct population highlights \textit{secular migration} as a potentially interesting component of NSC assembly.

Finally, while previous studies of dwarf-galaxy NSCs generally find a significant contribution from \textit{star cluster} infall, this channel is negligible in our sample, contributing less than $3\%$ of the NSC mass in all cases. This is particularly striking given the substantial population of star cluster-like structures that forms throughout our simulations. Thus, the lack of cluster-delivered mass does not reflect an inability of LYRA to form such objects, but rather indicates that most clusters do not migrate into the NSC. We nevertheless regard these fractions as lower limits, as some clusters may be missed by \texttt{subfind} or artificially disrupted by numerical effects before reaching the NSC. Any additional contribution from this channel would necessarily reduce the inferred \textit{merger} and/or \textit{secular migration} fractions, while leaving the \textit{NSC-Born} component unaffected, as the latter is defined independently of \texttt{subfind} associations.

\begin{figure*}
    \centering
    \includegraphics[width=0.98\linewidth]{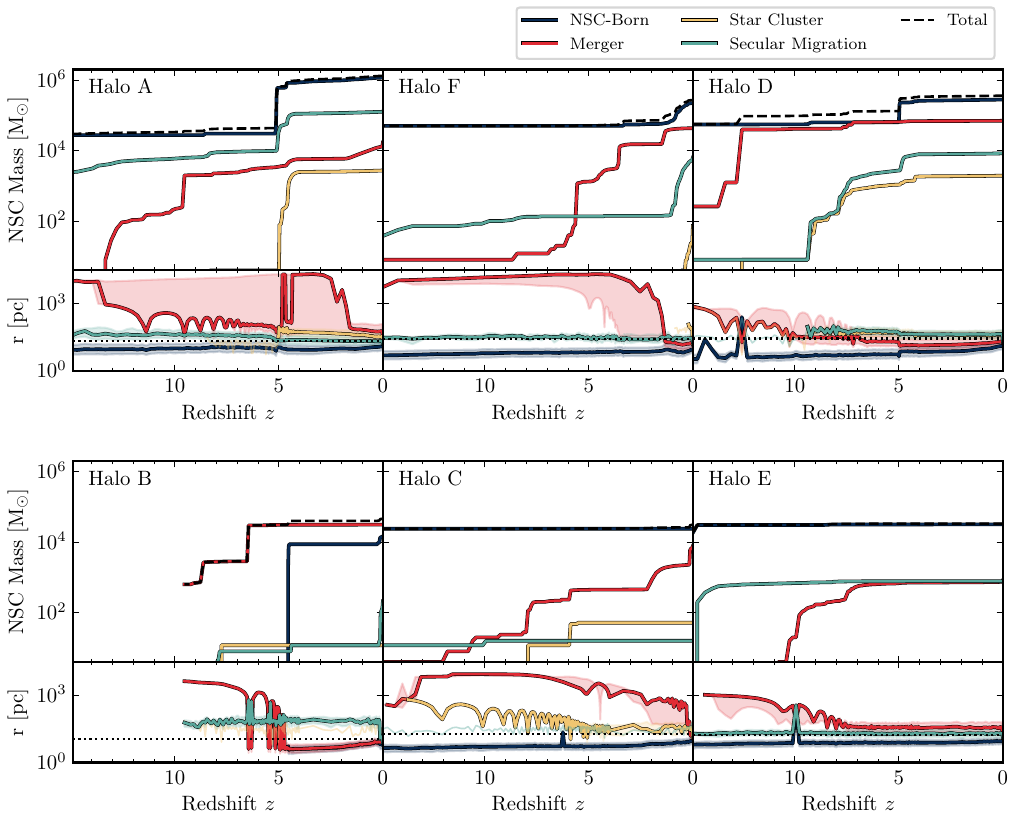}
    \caption{Nuclear star cluster mass growth separated by the different contributions for each of the LYRA haloes. The total NSC mass is shown as a dashed line in each panel. Each bottom panel shows the median distance to the NSC centre for each category of stars. The shaded regions show the interquartile range of the distribution of particles.}
    \label{fig:mgrowth}
\end{figure*}

To gain insight into the emergence of the NSCs over time, we explore their mass assembly separated into their respective origin channels. In each of the two panels of Fig.~\ref{fig:mgrowth}, we present the mass evolution on the top, and the median distance of the particles to the NSC centre on the bottom, split into origin category and for all LYRA galaxies. In all cases except Halo~B, we see that the NSC mass assembly is dominated by the \textit{NSC-Born} component already at early times ($z\geq 10$). This confirms what we discussed about the emergence of NSCs in LYRA (see Section~\ref{sec: emergence}): the NSC component appears to form first and the galaxy continues to accrete and form stars around it.

In addition, as expected from Fig.~\ref{fig:Fractions}, we see how the \textit{merger} component adds to the NSC mass in sharp increases, corresponding to different mergers. Note that the mass here is added to the corresponding category the first time it crosses $r_\mathrm{NSC}$. While mass can sharply increase, we might see particles orbiting around the NSC for a while before merging completely (see, for instance, the bottom panel for Halo A). Regarding the \textit{star cluster} contribution, we see that in most cases this occurs at later times. Nevertheless, we remind the reader that this component only makes up a small fraction of the final NSC mass, in many cases, contributing fewer than 100 star particles. Therefore, we need to interpret their implications with caution. 

In terms of the \textit{secular migration} contribution, we see that the mass growth in this component correlates closely with the increase in \textit{NSC-Born} mass. The clearest examples are the increase at $z\sim5$ for Haloes A and D, as well as the increase at $z\sim1.5$ for Halo F. The reason for this is that when the \textit{in-situ} star formation occurs in the NSC, there is also star formation close to the NSC radius. These stars end up crossing the NSC radius a short time thereafter. This star formation can also occur in gravitationally bound star clusters near the NSC, which can quickly infall into the NSC \citep{Poulain:2025}. However, these star clusters are not massive enough and too short-lived to be flagged as star cluster candidates with our methodology (Sec.~\ref{sec:particle-label}), and are therefore associated with the \textit{secular migration} channel. As a result, the \textit{secular migration} contribution should be interpreted with these limitations in mind, being relevant to capture potentially under-represented mass from the \textit{star cluster} channel.

The case of the NSC in Halo B deserves particular attention. It is dominated by a \textit{merger} contribution that accounts for over $60\%$ of the NSC mass, mostly from a single progenitor. This NSC shows no \textit{in-situ} contribution before $z=5$. This could be because we define the main branch of the merger tree based on the DM halo, which does not necessarily track the progenitor with the most stellar mass. Additionally, at $z\sim10$, this halo undergoes a significant interaction with another galaxy of similar halo mass, i.e. a near 1:1 merger. This merger likely disturbs the centre of the halo, causing any subsequent star formation to occur slightly outside $r_\mathrm{NSC}$, as shown by the existing \textit{secular migration} contribution. Once the merger disruption settles, the onset of reionization suppresses further star formation, delaying \textit{in-situ} growth, as suggested by the $z=5$ starburst shown in Fig.~\ref{fig:mgrowth} \citep[see also][]{Sureda:2026}. Selecting a different progenitor as the main branch could imply reclassifying some of the earliest accreted mass as \textit{in-situ}, though we have not verified whether it formed at the centre of the accreted galaxy. Nevertheless, these early accretion episodes account for only a small fraction of the merger mass in this halo; therefore the picture of a merger-dominated NSC would remain. This then points to a different NSC formation scenario in this halo than the one discussed in Sec.~\ref{sec: emergence}.

\subsection{Properties of the different populations}

\subsubsection{Mass distribution}\label{sec:mass_dist}

\begin{figure}
    \centering
    \includegraphics[width=0.95\linewidth]{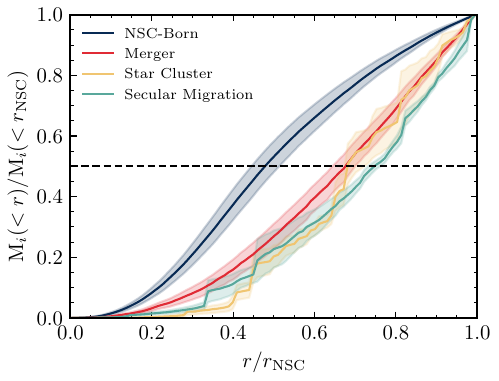}
    \caption{Cumulative radial mass distribution in the NSCs for each stellar component averaged across all LYRA haloes. The solid line shows the average while the shaded regions show the standard deviation in each radial bin. This figure highlights how concentrated the \textit{NSC-Born} component is compared with the rest of the stellar components, reaching $50\%$ of its mass within half of the NSC radius, as opposed to $\sim0.7\,r_\mathrm{NSC}$ for the others.}
    \label{fig:massprofile}
\end{figure}

In order to get a sense of how the different categories of stars are distributed in the NSC, we compute their cumulative radial mass distributions. In Fig.~\ref{fig:massprofile} we see these distributions averaged across all haloes. We notice that the \textit{NSC-Born} component is the most centrally concentrated component in all haloes, with the radius enclosing half of the NSC mass, $r_{50} < 0.5\,r_\mathrm{NSC}$. The other components are generally distributed in larger radii, with $r_{50} $ within $ (0.6 -  0.8)\,r_\mathrm{NSC}$, even considering the scatter across haloes. It is worth highlighting that the \textit{secular migration} category is the least concentrated with a significantly smaller scatter at larger radii. 

\begin{figure}
    \centering
    \includegraphics[width=\linewidth]{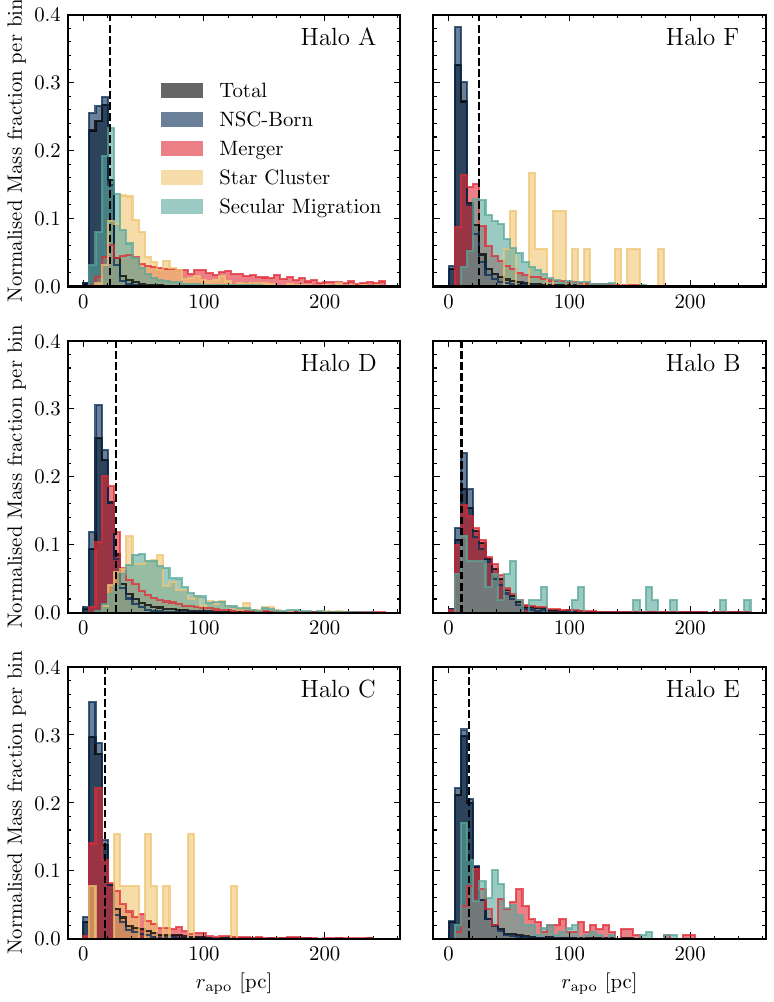}
    \caption{Normalised apocentre distribution of NSC stars at $z=0$, separated into the different growth channels. Each individual histogram is normalised to its individual mass contribution, i.e., each histogram individually adds up to 1. The vertical dashed line shows $r_\mathrm{NSC}$ for each halo. This highlights how centrally-concentrated the \textit{NSC-Born} component usually is compared to the rest. While their orbits usually have a peak close to $r_\mathrm{NSC}$, they do have more extended tails with larger apocentres.}
    \label{fig:apocentres}
\end{figure}

The results presented so far are based on a sharp radial cut to define the NSC region. This is in principle not ideal as in reality this region could be contaminated by stars on highly eccentric orbits or stars passing by. However, we made this choice to allow for a fairer comparison with observations, where it is extremely difficult to remove these contaminants without a detailed dynamical modelling of the region \citep{Fritz:2016, Vasilev:2026}. Hence, to assess the impact of this strict radial definition on our results, we compute the apocentre distributions of NSC stars and explore how the mass is distributed as a function of apocentre distance. To achieve this, we use \texttt{AGAMA} \citep{Vasukev:2019_AGAMA} to compute the gravitational potential and infer the apocentre of the individual particles within the NSC based on their current positions and velocities. For each halo, we compute the apocentre distribution of each category and show it in Fig.~\ref{fig:apocentres}. Each particle is weighted by its mass and each distribution is normalised by its contribution to the total NSC mass, such that they all add up to 1. From this figure, we can see that, except for Halo B, the \textit{NSC-Born} component is in all cases the most centrally concentrated and contains most of this contribution mass within $r_\mathrm{NSC}$ (shown as a vertical dashed line), with a small tail in the distribution that quickly reaches 0 within $\sim 80$ pc. Interestingly, in Halo B, this component is less concentrated and has a wider distribution, resembling more the one of the \textit{Merger} component, although with slightly more concentrated orbits. This is consistent with what we discussed in Sec.~\ref{sec:OriginOfNSCStars}, supporting the idea of a merger-driven NSC formation for this particular halo. 

For the other haloes, however, the apocentre distribution for the  merger component contains a considerable fraction of small apocentres, with a significant tail of orbits with larger apocentres. Although one could expect merger-contributed stars to sit in larger orbits, most of the mergers that contribute significantly to these NSCs occur at early times ($z > 3$ or more than 11 Gyrs ago) and have therefore had enough time to virialise their orbits and, for the case of these stars, sink to the centre of the galaxy.

A similar behaviour might be expected from the \textit{star cluster} contribution. In this case, these objects have slightly larger apocentres in general, reflecting their later infall times, as we can also infer from the mass growth shown in Fig.~\ref{fig:mgrowth}. Nevertheless, the statistics of this population are poorer given the smaller number of particles associated with this channel, explaining the high noise in the distribution for some cases, e.g. Halo F and C. We also do not show this distribution for Halo B, given that this channel only contributes with 3 star particles.  

We additionally inspected the impact of selecting NSC particles based on their apocentre distances on the mass fractions for each category. As suggested by Fig.~\ref{fig:apocentres}, the majority of \textit{NSC-Born} stars have apocentres within $r_\mathrm{NSC}$, and since they are the dominant contribution, the overall mass fractions are mostly unaffected by this selection criterion. In general, excluding particles with $r_\mathrm{apo}>r_\mathrm{NSC}$ removes mass from the \textit{Secular Migration} and \textit{Merger} channels, resulting in a median reduction of $63\%$ and $35\%$ in their mass contributions, respectively. Since the contribution from these two channels was already small, this only results in a median increase of $10\%$ in the mass fraction of \textit{NSC-Born} stars.

Finally, we also explore how the stellar kinematics shown in Fig.~\ref{fig:vdisp} vary if we isolate different components for the NSC region. We focused on Halo A which is the only halo with a rotation signature in the NSC region. We found that removing the \textit{NSC-Born} component from the sample erases the enhancement of the rotational velocity in the NSC region, preserving the same rotation level as the rest of the region shown in the figure. This indicates that while there is an overall rotation signature from the central region of the galaxy, the enhancement displayed within the NSC region is driven by the \textit{NSC-Born} stars. If we also remove the \textit{secular migration} stars, the transition at $x\sim0$ is even smoother, suggesting that these can also contribute to this signal to a lesser degree. This is reasonable considering these stars are usually formed in the vicinity of the NSC region. The reason for Halo A being the only one displaying this rotation signature can be related to being both the most massive galaxy and NSC within our sample; with a small sample, however, it is hard to be definitive about this.

In general, we see complex structures among the different components that make up the NSCs. While we can quickly identify the \textit{NSC-Born} component due to its compactness, it is not trivial to disentangle the rest of the components from each other. This is especially true if one were to consider observations of unresolved NSCs in dwarfs, where it becomes extremely difficult to infer the dynamics of the NSCs or even resolve their stellar kinematics in a reliable way. This emphasises the need for other signatures that can act as tracers for these different components.

\subsubsection{Age and Metallicity}

\begin{figure}
    \centering
    \includegraphics[width=\linewidth]{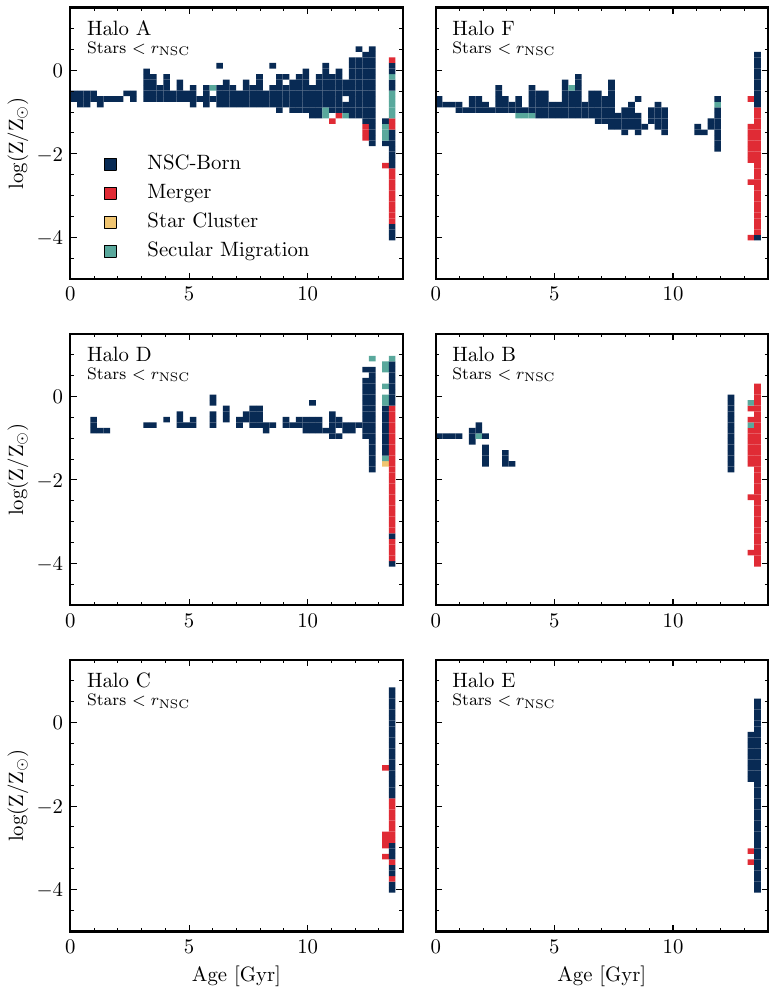}
    \caption{Age-metallicity distribution for each of the LYRA NSCs. Each bin is colour-coded according to its dominant channel in terms of mass.}
    \label{fig:agemet}
\end{figure}

Although the physical distributions of NSCs are informative, the most relevant information we can retrieve that is useful for making predictions for observations of unresolved systems combines the ages and metallicity of these populations. Since these form through different mechanisms, one would expect to find differences in this plane. We briefly mentioned that the \textit{NSC-Born} population is dominated by old stars in all galaxies. However, depending on their corresponding star formation histories, we can still expect a large age spread in this population. In Figure~\ref{fig:agemet} we show the age-metallicity distribution of the NSC stars for all haloes. For each bin in this plane, we compute the mass contribution of each population and use this information to label each bin with the most significant contributor. As expected, the tail at younger ages is dominated by the \textit{NSC-Born} component, with the youngest population ($\lesssim 5\,\mathrm{Gyrs}$) displaying a relatively small spread in metallicity compared to the oldest stellar populations. However, once we focus on the oldest bins, there is a lot of degeneracy among the populations, with the merger contribution dominating the low-metallicity tail in most cases but with a non-negligible contribution from the other components. Therefore, while it is possible to discern the origin of some of these stars from the stellar ages, it would require resolving ages with extremely high precision, which is currently unfeasible, especially at these old stellar ages. 

The other option would be to use specific abundances, which are currently more reliable, to recover some of this information. For instance, one could look for signatures including the expected abundance anti-correlations in globular clusters, such as between $\left[\mathrm{Al}/\mathrm{Fe}\right]$ and $\left[\mathrm{Mg}/\mathrm{Fe}\right]$, or between $\left[\mathrm{Na}/\mathrm{Fe}\right]$ and $\left[\mathrm{O}/\mathrm{Fe}\right]$ \citep{Mucciarelli:2009,Bastian:2018} in order to assess the importance of star clusters in the old stellar population. 

\section{Discussion}\label{sec:discussion}

\subsection{Expanding our understanding of nuclear star cluster formation}

Nuclear star cluster formation is still a topic under debate in the current literature. However, it is mostly studied through the two key formation channels already discussed here: globular cluster infall and in-situ star formation \citep{Neumayer:2020,Fahrion:2021}. We have addressed this topic with the aim of understanding and expanding our view of the different formation and growth channels for NSCs. 

Our sample of simulated galaxies covers the dwarf galaxy regime, in which observations suggest that the main formation channel for NSCs in these galaxies is the infall of globular clusters \citep{Fahrion:2022}. Our findings show a NSC growth dominated by an in-situ star formation component (our \textit{NSC-Born} component) in all galaxies except for Halo B. However, the corresponding  stellar ages indicate this in-situ star formation occurred in the early stages of galaxy formation. This explains why we have NSCs with a high fraction of old metal-poor stars, in line with observations in this mass regime, that also, counter-intuitively have a large fraction of mass in \textit{NSC-Born} stars. 

Nevertheless, the results presented in this work highlight the complexity of NSC formation and show how different growth channels are at play. Of particular interest is the contribution from galaxy mergers that contributes significantly and can even dominate the mass growth of NSCs as shown in the case of Halo B. In addition, other works have shown the impact of galaxy mergers in triggering the formation of clustered star formation giving rise to the emergence of an NSC \citep{Gray:2025}. This emphasises the complexity of galaxy mergers that may play a larger role than previously expected in the context of NSC formation and growth. 

\subsection{Caveats and Outlook}\label{sec:caveats}

As with any galaxy formation model, LYRA necessarily relies on a number of assumptions and approximations. There are, in particular, several physical processes that are not yet included that could impact some of the results presented in this work. For instance, the model currently does not include an implementation for SN-Ia and, therefore, does not account for their contribution to chemical enrichment. This is particularly important when considering individual element abundances with respect to iron, such as $\left[\alpha/\mathrm{Fe}\right]$ ratios. For this reason, we restrict our analysis of the NSC chemistry to overall stellar metallicities, which provide a robust measure of the global metal content without relying on the detailed abundance patterns affected by SNe-Ia.

The current model also does not include pre-supernova feedback from massive stars. This process can affect star formation in dense molecular clouds, as radiation and stellar winds from massive stars can heat and disrupt the gas in which they form before the onset of the first supernovae. Its absence could therefore allow star formation to proceed more efficiently in some dense environments than it would with this additional feedback channel. This is potentially relevant for the detailed star-formation histories of the galaxies and NSCs considered here and may contribute to some of the differences between our simulated systems and observations.

The missing radiation from massive stars may have the strongest impact on the detailed evolution of the stellar component. However, several of the main conclusions of this work are expected to be more robust. In particular, the non-negligible contribution from galaxy mergers would remain unaffected, as well as the overall conclusions for the other channels. Indeed, the absence of pre-supernova feedback could explain why our entire sample is nucleated, even if the nucleation fraction at these mass scales is typically below $10\%$ \citep{Sanchez-Janssen:2019, Hoyer:2021}, although these estimates are not for isolated dwarfs. 

A separate limitation concerns the treatment of the star particles in LYRA, which are modelled as collisionless systems. While this is not an issue at the scale of the galaxy, it breaks down at star cluster scales, where two-body relaxation is expected to be relevant over a Hubble time. Since this effect is not captured, the mass loss of star clusters is slowed down, artificially enhancing their survivability \citep[see][]{Gutcke:2024}. Possible avenues to incorporate such effects would rely on implementing  similar approaches to those presented in \citet{Rantala:2024, Lahen:2025}. Nevertheless, this limitation does not affect the results presented in this work because the star cluster contribution is already negligible.

Along the same lines, the collisionless treatment also implies that some of the NSCs found in LYRA are only marginally resolved. In particular, the NSC in Halo B has a size $r_\mathrm{NSC}<3 \epsilon$, where $\epsilon=4\,\mathrm{pc}$ is the softening length, falling under the softening scale. While this could impact the kinematic measurements within the NSC for this halo, it would not affect the conclusions regarding the origin of its stars. Furthermore, this could influence the nucleation fraction in our sample, which would be an important factor to consider for a larger sample. In contrast, the NSCs in Halos A, D, and especially F (where $\epsilon=1\,\mathrm{pc}$) are better resolved, highlighting that these emerge physically from the simulation and are less subject to numerical noise. 

With future updates to the model, we expect to revisit some of these results. In particular, including the additional feedback and chemical-enrichment processes discussed above will allow us to test if the nucleation fraction changes and, eventually, whether detailed chemical abundance patterns can be used to reconstruct the assembly histories of NSCs from observable quantities. Nevertheless, the more immediate future work will focus on expanding the sample of simulated dwarfs, incorporating the runs presented in \citet{Brown:2026}. This will allow to also explore galaxies in lower mass regimes with a larger sample to increase our statistics and will be extremely relevant for making predictions regarding the fraction of nucleated dwarfs, as previously mentioned. Furthermore, the runs employ different prescriptions for the $\mathrm{H}_2$ dissociating Lyman-Werner (LW) background radiation at high redshift, allowing us to explore the impact of this prescription on NSC formation. This could have a potential impact on the contribution from galaxy mergers as some these galaxies can be entirely dark (i.e. without stars) depending on the prescription used, resulting in an non-existent contribution to the NSC mass. In this context, \citet{Gray:2025} discusses how changes in the background radiation field that suppress star formation in this mass regime could shift the mass scale for NSC formation via merger-driven starbursts, a process already discussed in other works \citep{Springel:2005, Lelli:2014, Zhang:2020}. While we do not report any NSC formed through this mechanism (i.e. merger-driven starburst), it is also worth investigating in future work.

\section{Conclusions}\label{sec:conclusions}

In this work we investigated the growth of nuclear star clusters in dwarf galaxies using the high-resolution LYRA simulations. We identify NSCs in all six galaxies in our sample, with stellar masses between $M_\mathrm{NSC}=3.24\times10^4 - 1.32\times10^{6} \, \msun$, consistent with observed NSCs. We explore the emergence and the main growth channels of these NSCs, expanding on the classical channels, i.e., in-situ star formation within the NSC and star cluster infall. We also consider the contribution from galaxy mergers and secular migration of stars born within the host galaxy (see Section~\ref{sec:particle-label}). We explore the different NSC properties and focus on reconstructing the assembly history of the NSCs in terms of these different growth channels. The key findings of this work are as follows:

\begin{enumerate}
    \item The identified NSCs in LYRA are broadly consistent with the $M_\mathrm{NSC}-M_\star$ relation derived from observations (see Section~\ref{sec:NSC-properties}). The three most massive galaxies in our sample overlap with the stellar mass regimes explored in the literature, and are consistent various datasets and the relation within its scatter. The smaller galaxies extend below the mass ranges covered by current observations. These galaxies fall systematically below the relation extrapolated to these masses but are still within its expected scatter.
    \item The emergence of NSCs in the LYRA simulations occurs within the first Gyr of the evolution history of our galaxies (see Section~\ref{sec: emergence}). The mechanism driving this emergence is that after a compact stellar component forms within the main halo, it both accretes and continues to form stars in a more extended region, producing the double component in the surface density profile, characteristic of the presence of a NSC.
    \item Most of the NSC mass is formed within the NSC region (see Fig.~\ref{fig:Fractions}), dominating the mass budget in 5 of our 6 haloes. We stress that most of this contribution is made up of old stars ($\mathrm{Age}>2$ Gyrs), i.e. the \textit{NSC-Born} (or in-situ) component is already in place at early times.
    \item There is a non-negligible mass contribution from stars formed in accreted systems, even dominating the mass budget in one of our haloes. Most of this contribution is made by a single merger at early times. This is significant for cases where the NSC host galaxy is quenched and can only grow through mergers.
    \item Contrary to expectations for this mass regime, stars contributed via the infall and disruption of star clusters are negligible in our sample, accounting for less than $3\%$ of the NSC mass. This happens despite the identification of a non-negligible population of star cluster candidates in each run. We stress that the modelling of star clusters in the current model is subject to numerical uncertainties (Sec.~\ref{sec:caveats}), although, this should not impact the key results presented here. Still, while it can be virtually impossible to distinguish stars from this channel from those in other channels, this result highlights the importance of exploring multiple growth channels in combination with those already established in the literature.
    \item When present, the younger stellar population in the NSCs is associated with the \textit{NSC-Born} channel. In contrast, the old stellar population originates from various channels, including the \textit{NSC-Born} one (see Fig.~\ref{fig:agemet}), complicating the identification of different growth channels in the old population based on age estimates alone. While the chemistry can leave potential signatures to use, metallicity alone is not enough to distinguish among the different populations, as these can reside in the same metallicity range, including the old component of the \textit{NSC-Born} stars. This highlights the need for detailed chemical information in order to disentangle these old stellar populations.
    \item While for some galaxies we reproduce the expected high fraction of old metal poor stars, suggesting a dominant GC accretion growth, when exploring the origin of those stars, we find that these NSCs have a dominant contribution from \textit{NSC-Born} and \textit{Merger} stars, highlighting that multiple channels (including \textit{in-situ} formation) can reproduce the signatures that we may otherwise expect to be a GC infall dominated NSC.
    \item Finally, even though these NSCs are dominated by a \textit{NSC-Born} component, only one of the haloes (Halo A) displays stellar kinematics expected for this scenario, i.e., rotation signatures leftover from the accreted gas dynamics. Nevertheless, in all other haloes, the NSCs appear as a kinematically cold but non-rotating self-gravitating component. Moreover, if we exclude the \textit{NSC-Born} stars from the analysis, only Halo A shows a significant difference, where the slight $v_\mathrm{los}$ enhancement in the NSC region vanishes and only the overall galaxy rotation signature remains, suggesting this excess is associated with these \textit{NSC-Born} stars. The fact that only Halo A displays this rotation signature may be related to Halo A being the most massive NSC and galaxy in our sample.
\end{enumerate}

While we discussed some of the caveats of the LYRA model (see Section~\ref{sec:caveats}) and their possible impact on the results presented here, we emphasise the importance of exploring different growth channels for NSCs. The various channels contributing to the NSC mass discussed in this work suggest a complex assembly history for NSCs. We found some mild separation of these growth channels in their concentration, as well as the apocentre distributions and age. However the overall metallicities do not contribute meaningfully in distinguishing between these channels. Further improvements to the treatment of the detailed chemistry in the simulations (for instance tracking $\left[\alpha/\mathrm{Fe}\right]$) are needed to enable more direct comparison with expectations for globular clusters contributing to NSC growth. Although we are able to reconstruct this assembly history from the simulation data, it is important to place this into perspective to assist the interpretation of current and future observations of NSCs, and to inform other relevant processes operating near the centres of galaxies at these mass scales.

\section*{Acknowledgements}

The authors would like to thank Dr John Helly for their support with the \texttt{D-halos} tree code. JS acknowledges support from the Science and Technologies Facilities Council (STFC) through the studentship grant ST/X508354/1, and the support from an ESO Studentship (2024/2025). AF is supported by a Sweden's Wallenberg Academy Fellowship. SB is supported by the UKRI Future Leaders Fellowship (grant numbers MR/V023381/1 and UKRI2044). JED is supported by the United Kingdom Research and Innovation (UKRI) Future Leaders Fellowship `Using Cosmic Beasts to uncover the Nature of Dark Matter' (grant number MR/X006069/1). KF acknowledges funding from the European Union’s Horizon 2020 research and innovation programme under the Marie Sk\l{}odowska-Curie grant agreement No 101103830.
We acknowledge the computing time provided by the Leibniz Rechenzentrum (LRZ) of the Bayrische Akademie der Wissenschaften on the machine SuperMUC-NG (pn73we). This work used the DiRAC@Durham facility managed by the Institute for Computational Cosmology on behalf of the STFC DiRAC HPC Facility (www.dirac.ac.uk). The equipment was funded by BEIS capital funding via STFC capital grants ST/K00042X/1, ST/P002293/1, ST/R002371/1 and ST/S002502/1, Durham University and STFC operations grant ST/R000832/1. DiRAC is part of the National e-Infrastructure.

\section*{Data Availability}

The data presented here is available upon reasonable request to the corresponding author.



\bibliographystyle{mnras}
\bibliography{references} 




\appendix

\section{NSC Emergence in all haloes}\label{app:emergence_all}

\begin{figure}
    \centering
    \includegraphics[width=\linewidth]{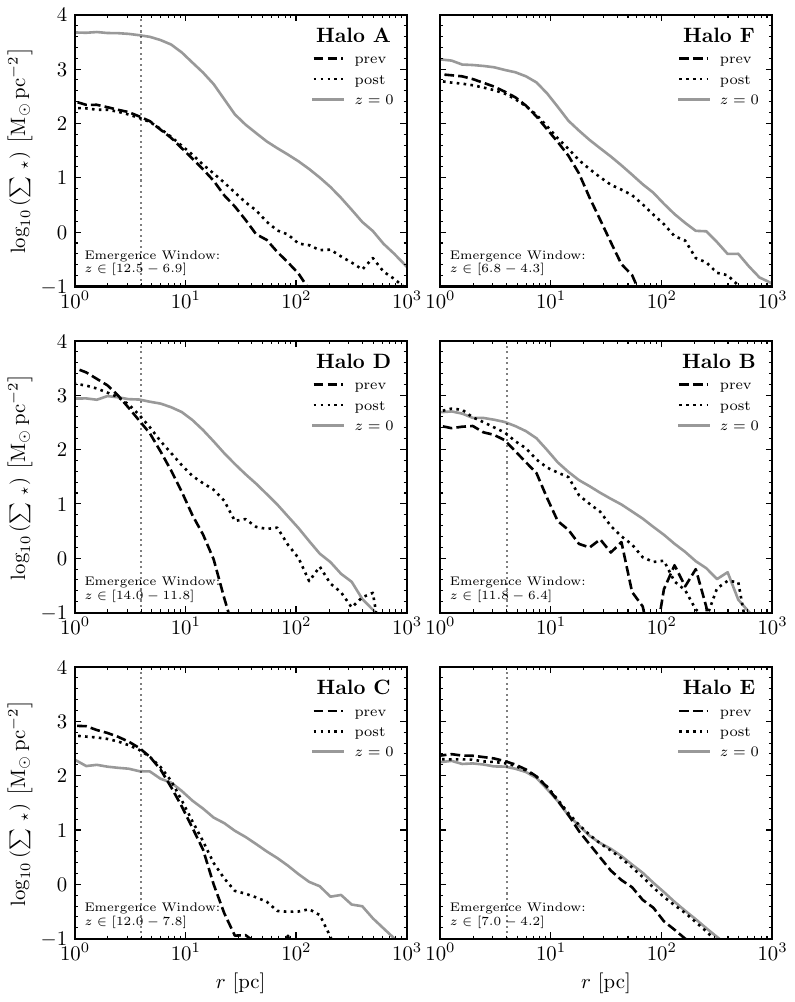}
    \caption{Median surface density profiles (as in Fig.~\ref{fig:emergence}), for all LYRA haloes. Dotted lines show the median profiles after the NSC emergence, while dashed lines show the median profiles before. The light solid line shows the $z=0$ profile for comparison. The dotted vertical lines indicate the softening length for each run. In Halo F this is $1$ pc.}
    \label{fig:app:emergence}
\end{figure}

For completeness, in the same way as for Halo F in Fig.~\ref{fig:emergence}, we show the comparison between the median surface density profiles of all halos in snapshots previous and after the emergence of the NSCs in Fig.~\ref{fig:app:emergence}. 

We note that the surface density profiles for haloes D and C shown in this figure could appear to be modelled by a single component. However, the extension of the profile requires the second component and is also preferred by our AIC criterion. This is also shown clearly in Fig.~\ref{fig: NSC surface density profile}. Moreover, the $z=0$ profiles of these two haloes show a decrease in the amplitude towards the centre, in comparison with the median profiles around the emergence window. We attribute this to the effect of softening in this regions (see the dotted vertical lines) as we can also see a flattening of the surface density within this region. However, this does not change our interpretation of the emergence of the NSCs since relaxation requires longer timescales than the emergence window, and we see the NSCs in place at quite early times.

\bsp	
\label{lastpage}
\end{document}